\documentclass[aps,prl,reprint,10pt,twocolumn,showpacs]{revtex4-1}
\usepackage{graphicx}
\usepackage{amssymb}
\usepackage{amsmath}
\usepackage{amsfonts}
\usepackage{color}
\usepackage[latin1]{inputenc}
\usepackage{float}
\usepackage{comment}
\usepackage{subfigure}

\newcommand{\lrangle}[1]{\langle{#1}\rangle}

\newcommand{\vinf}{v_\infty}

\begin{document}
\title{Universality of fluctuations in the Kardar-Parisi-Zhang class in high
dimensions and its upper critical dimension}
\author{S. G. Alves}
\author{T. J. Oliveira}
\author{S. C. Ferreira}
\affiliation{Departamento de F\'isica, Universidade Federal de Vi\c cosa,
36570-000, Vi\c cosa, MG, Brazil}
\date{\today}

\begin{abstract}
We show that the theoretical machinery developed for the Kardar-Parisi-Zhang
(KPZ) class in low dimensions are obeyed by the restricted solid-on-solid (RSOS)
model for substrates with dimensions up to $d=6$. Analyzing different
restriction conditions, we show that height distributions of the interface are
universal for all investigated dimensions. It means that fluctuations are not
negligible and, consequently, the system is still below the upper critical
dimension at $d=6$. The extrapolation of the data to dimensions $d\ge7$ predicts
that the upper critical dimension of the KPZ class is infinite.
\end{abstract}

\pacs{68.43.Hn, 68.35.Fx, 81.15.Aa, 05.40.-a}

\maketitle

Interface dynamics in nature is mostly a non-equilibrium process~\cite{barabasi}
and the Kardar-Parisi-Zhang (KPZ) universality class introduced by the
equation~\cite{KPZ}
 \begin{equation}
 \frac{\partial h (x,t)}{\partial t} = \nu \nabla^{2} h + \frac{\lambda}{2}
(\nabla h)^{2} + \xi,
\label{eqKPZ}
\end{equation}
where $\xi$ is a white noise of zero mean and amplitude $\sqrt{D}$, is certainly
one of the most relevant problems in non-equilibrium interface
science~\cite{krugrev,SasaSpohnJsat}.

Much is known about KPZ class in $d=1+1$ including exact
solutions~\cite{SasaSpo1,*Amir,*Calabrese,*Imamura}, experimental
realizations~\cite{TakeSano,*TakeuchiJSP12,*Yunker}, and fine-tuning properties
as universal schema for finite-time
corrections~\cite{Ferrari,TakeuchiJSP12,Oliveira12,Alves13} and for the
crossover to the stationary regime~\cite{TakeuchiCross,*HealyCross}. The
compilation of all results gave rise to the extended KPZ ansatz, {whose main ideas
were formerly introduced by Krug \textit{et al}.~\cite{KrugPRA92}}, for the
interface evolution in the non-stationary regime where the height at each surface
point evolves as
\begin{equation}
 h = \vinf t + s_\lambda(\Gamma t)^{\beta} \chi+\eta+\ldots,
\label{eq:htcorr}
\end{equation}
where  $s_\lambda=\mbox{sgn}(\lambda)$, $\beta$ is the growth exponent, and
$\chi$ is an stochastic variable, whose universal distribution depends
only on the growth geometry~\cite{PraSpo1}. {The spatial correlations,
which also depend on the growth geometry, are also known and given by the so-called Airy
processes~\cite{krugrev}.}
The parameters $\vinf$, $\Gamma$ and $\eta$ are non-universal,
being the last one responsible by a shift in the distribution
of the scaled height,
\begin{equation}
 q = \frac{h-\vinf t}{s_\lambda(\Gamma t)^{\beta}},
\label{eq:q}
\end{equation}
in relation to the asymptotic  distributions $\chi$. Except for the very specific
case where $\lrangle{\eta}=0$~\cite{Ferrari}, the shift vanishes as
$\lrangle{q}-\lrangle{\chi}\sim t^{-\beta}$.

Simulations of several models that, in principle, belong to the KPZ class
have shown that the ansatz given by Eq.~(\ref{eq:htcorr}) can be extended to
$d=2+1$ with universal and geometry-dependent stochastic
fluctuations~\cite{Healy,Alves12,Oliveira13,Healy2}. Despite of a possible
fragility of the KPZ equation to non-local perturbation discussed in
Ref.~\cite{Nicoli2013}, the universality for flat growth in $d=2+1$ was recently
observed in semiconductor~\cite{almeida2013} and organic~\cite{Healy2014} films.

The KPZ class in dimensions $d\ge 3$ is still an open problem with basic
unanswered questions. In particular, the upper critical dimension $d_u$ above
which fluctuations become irrelevant, independently of the strength of
non-linearity of the KPZ equation, is a controversial unsolved problem.
Several works based on mode-coupling theory
\cite{Cates93,*Doherty94,*Doherty95,*Colaiori01} and field theoretical approaches
\cite{Healy90,*Lassig97,*Fogedby06} suggest $2.8 \lesssim d_u \leq 4$,
whereas renormalization group
approaches~\cite{Canet2010,*Kloss2012,Castellano1998a,*Castellano1998b} and
simulations of KPZ models
~\cite{Pagnani2013,Schwartz2012,*Perlsman,Odor2010,Kim2013,Tu94,*Marinari02,Ala-Nissila,*Ala-Nissila2}
show that KPZ upper critical dimension, if it exists, is higher than four.
Particularly, in Refs. \cite{Castellano1998a,*Castellano1998b,Tu94,*Marinari02} was
suggested that $d_u = \infty$. {A short but comprehensive review of the
state of the art is presented in Ref.~\cite{Pagnani2013}.}

Much of  these discussions on $d_u$
were  based on scaling exponents. The squared interface width,
defined as the variance of the interface height profile, evolves according to
the {Family-Vicsek ansatz~\cite{FV}}
\begin{equation}
 W^2(L,t) \equiv \left\langle h^2 \right\rangle_c = L^{2\alpha}f\left(\frac{t}{L^{z}}\right),
\label{eq:rug}
\end{equation}
where $\lrangle{X^n}_c$ denotes the $n$th cumulant of $X$. The scaling function
$f(x)=x^{2\beta}$ for $x\ll 1$ and $f(x)=\mbox{const}$ if $x\gg 1$. The
roughness $\alpha$ and dynamical $z$ exponents obey the scaling relation
$\alpha+z=2$ independent of the dimension~\cite{krugrev}, {which was checked up to
5+1 dimensions~\cite{Odor2010}.}  For $1\ll t\ll L^z$,
$W\sim t^{\beta}$ where the growth exponent is given by $\beta=\alpha/z$. For $d
\geqslant d_u$, we have $\alpha=\beta=0$ and $z=2$.

In this work, we  investigate the classical restricted solid-on-solid (RSOS)
model {proposed by Kim and Korterlitz~\cite{KK}, which is widely accepted
as a prototype of the KPZ universality class~\cite{Pagnani2013}}, at the light
of the KPZ ansatz for interface fluctuations given by Eq.~(\ref{eq:htcorr}). In
the RSOS model, particles are deposited in random positions of an initially flat
substrate, represented by a $d$-dimensional lattice, given that the height
differences among nearest neighbors are not larger than a certain integer value
$m$, the height restriction parameter. We analyzed substrates with dimension up
to $d=6$ and show that in all dimensions the model follows the theoretical KPZ
machinery for interface fluctuations given by Eq.~(\ref{eq:htcorr}) as well as
growth velocity dependence on system size and substrate slope~\cite{krug90}.
Furthermore, the cumulant ratios and, consequently, the height distributions are
universal and depend only on $d$ for distinct height restriction parameters in
all investigated dimensions. It reads as fluctuations are not negligible and,
consequently, we are still below the critical dimension for $d=6$. The
distributions and corrections to the scaling are also presented.

It was shown that differences greater than unity in the height restriction
parameter of the RSOS model reduce finite-size and -time corrections and,
consequently, improves the scaling analysis~\cite{Kim2013}. So, we used  height
restriction $m=2$ to 8, depending on dimension. The case $m=1$ is hampered by an
initial layer-by-layer growth which makes the determination of universal
quantities very hard in high dimensions~\cite{Kim2013}. We considered
$d$-dimensional substrates of sizes up to $L_{max}$ and periodic boundary
conditions. Only results for the largest ones are shown. Sizes
$L_{max}=2^9,~2^7, 2^6, 2^5$ and number of independent samples
$N_s=2500,~1000,~250,~50$ were used for dimensions $d=3$ to 6, respectively. In
all cases, typically $\sim10^{11}$ surface points were used to compute
statistics of interface distributions.

\begin{figure}[hbt]
 \centering
 \includegraphics[width=8.5cm]{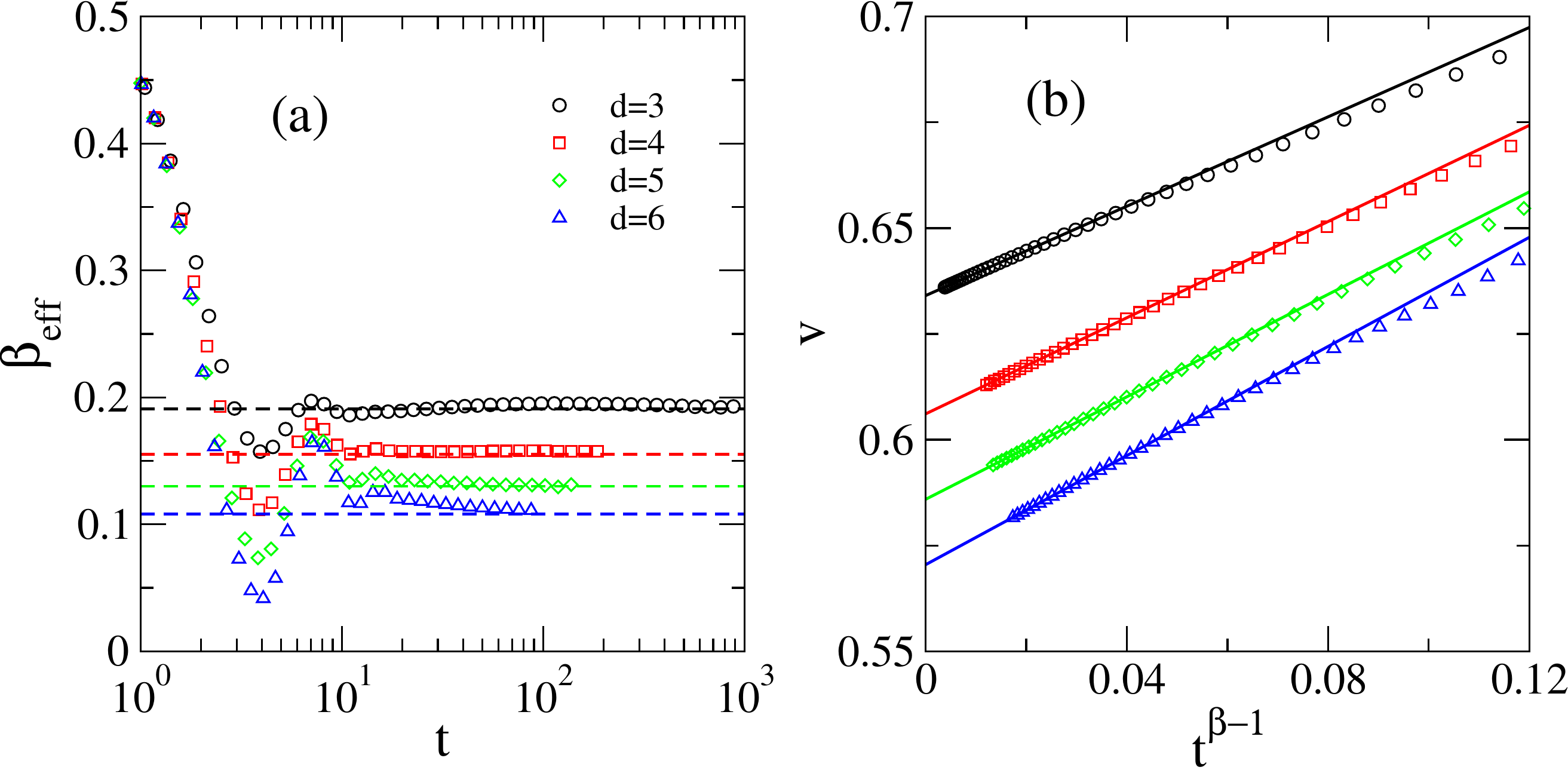}

 \includegraphics[width=8.5cm]{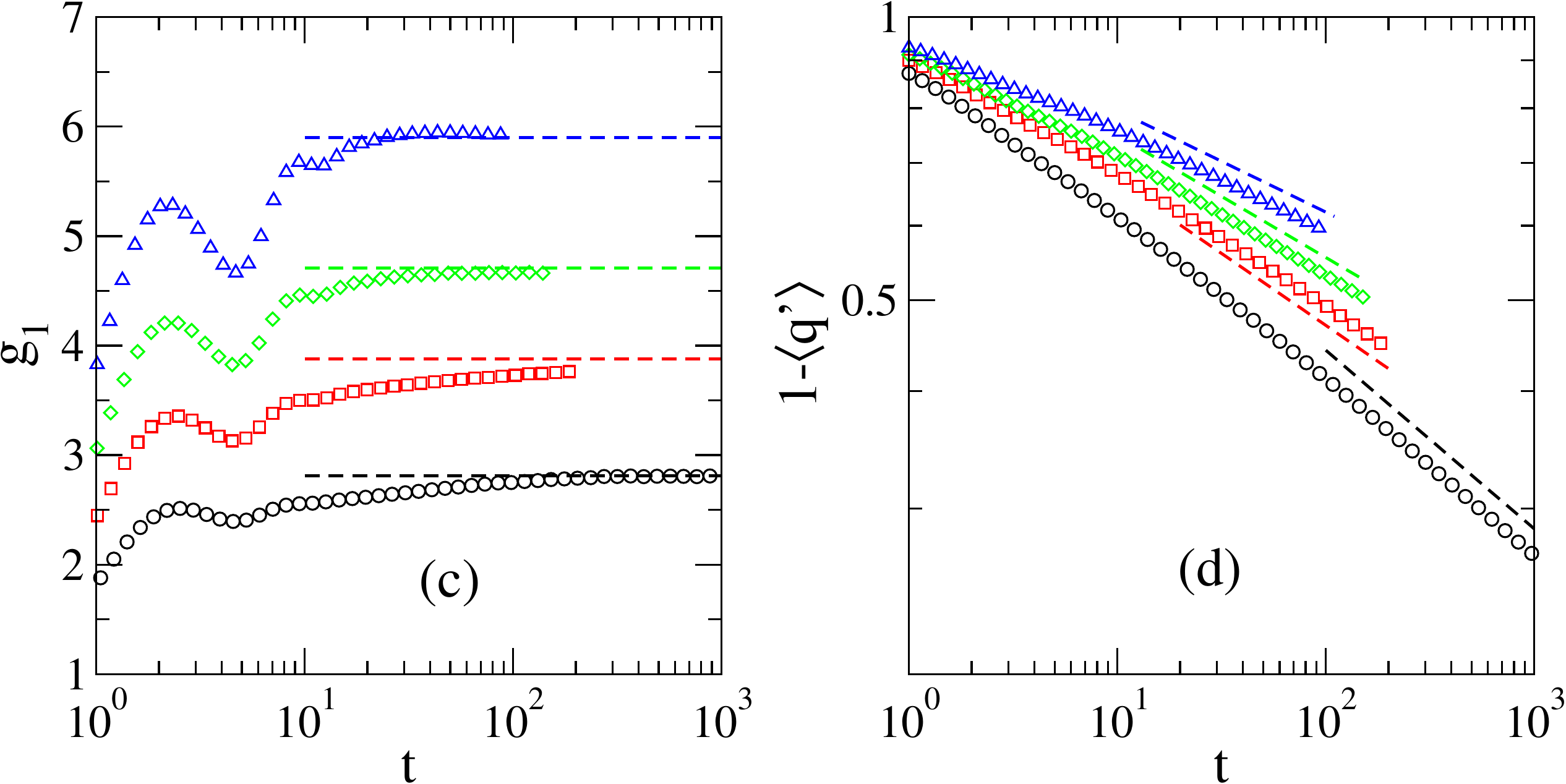}
 \caption{(a) Effective growth exponent defined as $d\ln W/d\ln t$ for RSOS
model in $d+1$ dimensions with height restriction $m=4$. Dashed lines are the
estimates for $\beta$. (b) Curves used to determine the asymptotic interface
velocity. Solid lines are linear regressions. (c) Determination of the quantity
$\Gamma^\beta\lrangle{\chi}$. Horizontal lines are the extrapolated values. (d)
Scaling of the shift in relation to the asymptotic distribution. Dashed lines are
power laws $t^{-\beta}$ using $\beta$ exponents given in Table~\ref{tab:ds}.}
 \label{fig:betav}
\end{figure}

According to Eqs. (\ref{eq:htcorr}) and (\ref{eq:rug}), the growth exponent can
be determined using the time dependence of $\lrangle{h^2}_c$.
Figure~\ref{fig:betav}(a) shows the effective growth exponent against time for
RSOS model in several dimensions. The exponents extrapolated to
$t\rightarrow\infty$ are show in Table~\ref{tab:ds}. The obtained values are in
very good agreement with the recentest results for $d\le
5$~\cite{Odor2010,Kim2013}, in particular for $d=4$, where most of the
discussion about the upper critical dimension has held on. Our estimate for
$d=6$ is consistent but more accurate than former simulations presented in
Ref.~\cite{Ala-Nissila,*Ala-Nissila2}. The interface velocity was calculated with
Eq.~(\ref{eq:htcorr}): The time derivative of $\lrangle{h}$ plotted against
$t^{\beta-1}$ is extrapolated via a linear regression~\cite{Alves13}. The
interface velocities  against $t^{\beta-1}$ for the RSOS model with $m=4$ are
shown  in Fig.~\ref{fig:betav}(b). Their asymptotic values are summarized in
Table~\ref{tab:ds}.

The shift of the distribution is investigated defining the quantities
$g_1=(\lrangle{h}_t-v_\infty)/\beta\rightarrow\Gamma^\beta\lrangle{\chi}$ and
$q'=(h-v_\infty t)/(s_\lambda g_1 t^{\beta})$~\cite{Alves13}.
Figure~\ref{fig:betav}(c) shows the curves $g_1(t)$ for RSOS model with $m=4$
and the corresponding extrapolated values. Equation~(\ref{eq:htcorr}) implies
that $1-\lrangle{q'}\simeq - (s_\lambda\lrangle{\eta}/g_1)t^{-\beta}$, which is
confirmed in Fig.~\ref{fig:betav}(d). Therefore, the KPZ ansatz
 including the correction $\eta$, Eq.~(\ref{eq:htcorr}), is very well obeyed by
RSOS model in $d=3-6$ dimensions, in analogy with the lower dimensional
cases~\cite{Oliveira12,Alves13,Oliveira13}. These results
simultaneously generalize the validity of the KPZ ansatz
and  confirm that RSOS model belongs to KPZ class
in higher dimensions.

\begin{figure}[!t]
 \centering
 \includegraphics[width=8.5cm]{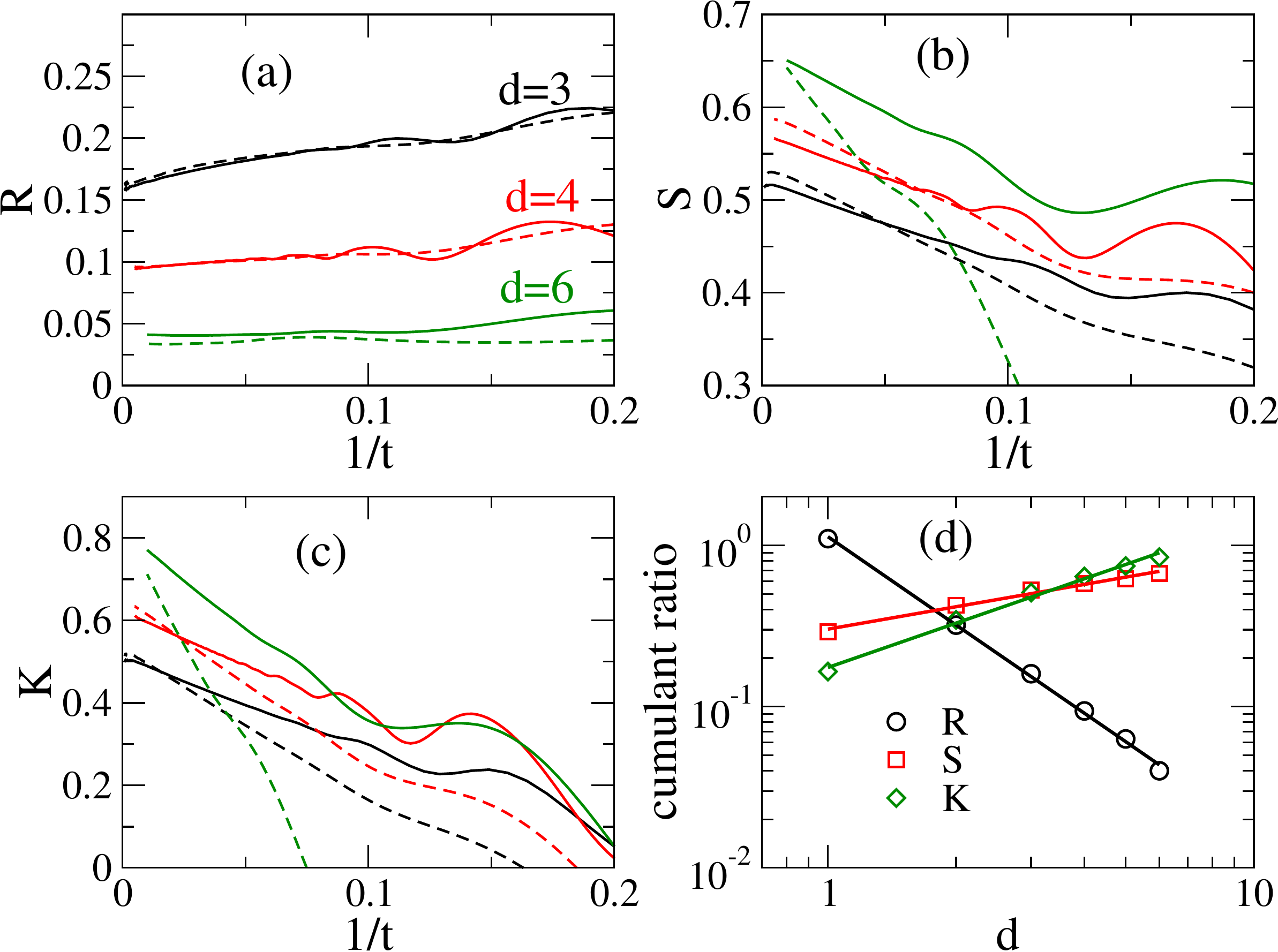}
 \caption{Dimensionless cumulant ratios  (a) $R=\lrangle{h^2}_c/\lrangle{h}^2$,
(b) $|S|=|\lrangle{h^3}|/\lrangle{h^2}_c^{3/2}$, and
(c) $K=\lrangle{h^4}_c/\lrangle{h^2}_c^2$ against inverse of time for RSOS model in
different dimensions indicated near the respective curves in panel (a).
The height restriction parameters of Table~\ref{tab:ds} are shown
for each dimension being the dashed lines the smaller ones.
Panel (d) shows the cumulant ratios as functions of the substrate dimension.
Data for $d=1$ and 2 were taken from Refs.~\cite{Alves13} and \cite{Oliveira13},
respectively.}
 \label{fig:RSK}
\end{figure}

Important properties of $\chi$ can be achieved through dimensionless
cumulant ratios. So, we define
quantities $g_n=\lrangle{h^n}_c/t^{n\beta}$ for $n\ge 2$ that, according to
Eq.~(\ref{eq:htcorr}), goes to $\Gamma^\beta\lrangle{\chi^n}_c$ for
$t\rightarrow\infty$~\cite{Alves13}. Therefore, the dimensionless ratios
$R=g_2/g_1^2$, $S=g_3/g_2^{3/2}$ (skewness), and $K=g_4/g_2^2$ (kurtosis)
 are independent of the particular model in the case of a universal
 $\chi$. Figures~\ref{fig:RSK}(a)-(c) shows the dimensionless cumulant ratios
against inverse of $t$, where one can see convergence to values that do
neatly vary with substrate dimension but does not with the height restriction
parameter. Extrapolated values are given in Table~\ref{tab:ds}.

\begin{table}[!b]
 \begin{tabular}{cccccc}\hline\hline
  Model & $\beta$ & $\vinf$ & $R$ & $S$ & $K$  \\ \hline
\multicolumn{6}{c}{$d=3$} \\  \hline
  RSOS ($m=2$) & 0.189  & 0.44650  &  0.156 & 0.53  & 0.50  \\
  RSOS ($m=4$) & 0.191  & 0.6340   &  0.163 & 0.53  & 0.52  \\\hline
\multicolumn{6}{c}{$d=4$} \\ \hline
  RSOS ($m=2$) & 0.150  & 0.41518  &  0.093 & 0.57  & 0.63  \\
  RSOS ($m=4$) & 0.155  & 0.6059   &  0.096 & 0.59  & 0.65  \\ \hline
\multicolumn{6}{c}{$d=5$} \\ \hline
  RSOS ($m=2$) & 0.13   & 0.39356  &  0.064 & 0.61  & 0.73  \\
  RSOS ($m=4$) & 0.13   & 0.5858   &  0.063 & 0.63  & 0.76  \\ \hline
\multicolumn{6}{c}{$d=6$} \\ \hline
  RSOS ($m=4$) & 0.11   & 0.57055  &  0.042 & 0.66   & 0.83  \\
  RSOS ($m=8$) & 0.10   & 0.7380   &  0.037 & 0.68   & 0.86  \\ \hline\hline
 \end{tabular}
\caption{\label{tab:ds} Numerical results for RSOS models in $d+1$ dimensions.}
\end{table}

Above $d_u$, where the surface is essentially flat, one  expects non-universal
cumulant ratios. For example, the linear Edwards-Wilkinson (EW)
equation~\cite{EW}, which is obtained with $\lambda=0$ in the KPZ equation, has
an exact solution and its upper critical dimension is $d_u=2$~\cite{barabasi}.
The numerical integration~\footnote{We used Euler method with time step $\Delta=0.01$} 
of the EW equation (with $\nu=2.5$ and $D = 0.5$) results in
kurtosis $K \approx -0.09$, $-0.18$ and $-0.26$ in $d=3, 4$ and $5$, respectively.
On the other hand, simulations of the random deposition with surface relaxation
(RDSR)~\cite{Family}, a typical discrete model in the EW class, provide $K
\approx 0.95$, $2.7$ and $4.3$ in the same dimensions. Notice that the up-down
symmetry of EW systems implies in a null skewness. The
kurtosis of EW equation becomes more negative with the substrate dimension
because the surface is smoothing for higher dimensions and, therefore, the
height distribution becomes narrower. Otherwise, for RDSR model the positive
increasing kurtosis is due to its discrete nature where only a very few
heights have non-negligible probabilities to occur.

In order to clarify the latter kurtosis dependence in
discrete systems, consider a toy surface with a height distribution
$P(\delta)=q$ if $\delta = - 1$, $P(\delta)=1-p-q$ if $\delta = 0$, $P(\delta)=p$
if $\delta = +1$, and $P(\delta)=0$ otherwise.
Both parameters $p$ and $q$ are very small mimicking the dependence on
dimension at $d>d_u$ and possibly  the
non-universal properties/parameters of the surface dynamics. One has
$\lrangle{\delta^n}=p-q$ for $n$ odd and $\lrangle{\delta^n}=p+q$ for $n$
even. To the leading order, the cumulants are, therefore,
$\lrangle{\delta^n}_c\simeq\lrangle{\delta^n}$. So, skewness and kurtosis are
$S\simeq (p-q)/(p+q)^{3/2}$ and $K\simeq 1/(p+q)$, respectively, which are clearly dependent
on the parameters $p$ and $q$. Moreover, the smother the interface
(smaller $p$ and $q$) the  larger $S$ and $K$. So,
the independence of RSOS model on the height restriction parameter is a
strong evidence that the model is still below $d_u$.

A smooth surface for $d\ge d_u$ also implies that the cumulant ratio $R$ is
null. Figure~\ref{fig:RSK}(d) shows the dependence of the cumulant ratios with
dimension. Our data corroborate the
discussion above: Increasing $d$ leads to smother surfaces and, thus, to
decreasing $R$ and increasing $S$ and $K$. An approximately power-law dependence on
dimensionality is found for all investigated ratios, being $R\sim d^{-1.8}$,
$S\sim d^{0.46}$ and $K\sim d^{0.92}$.
If these power laws hold for any
dimension, the KPZ class does not have an upper critical dimension, as
previously conjectured~\cite{Castellano1998a,*Castellano1998b,Tu94,*Marinari02}.

\begin{figure}[!t]
 \centering
 \includegraphics[width=8cm]{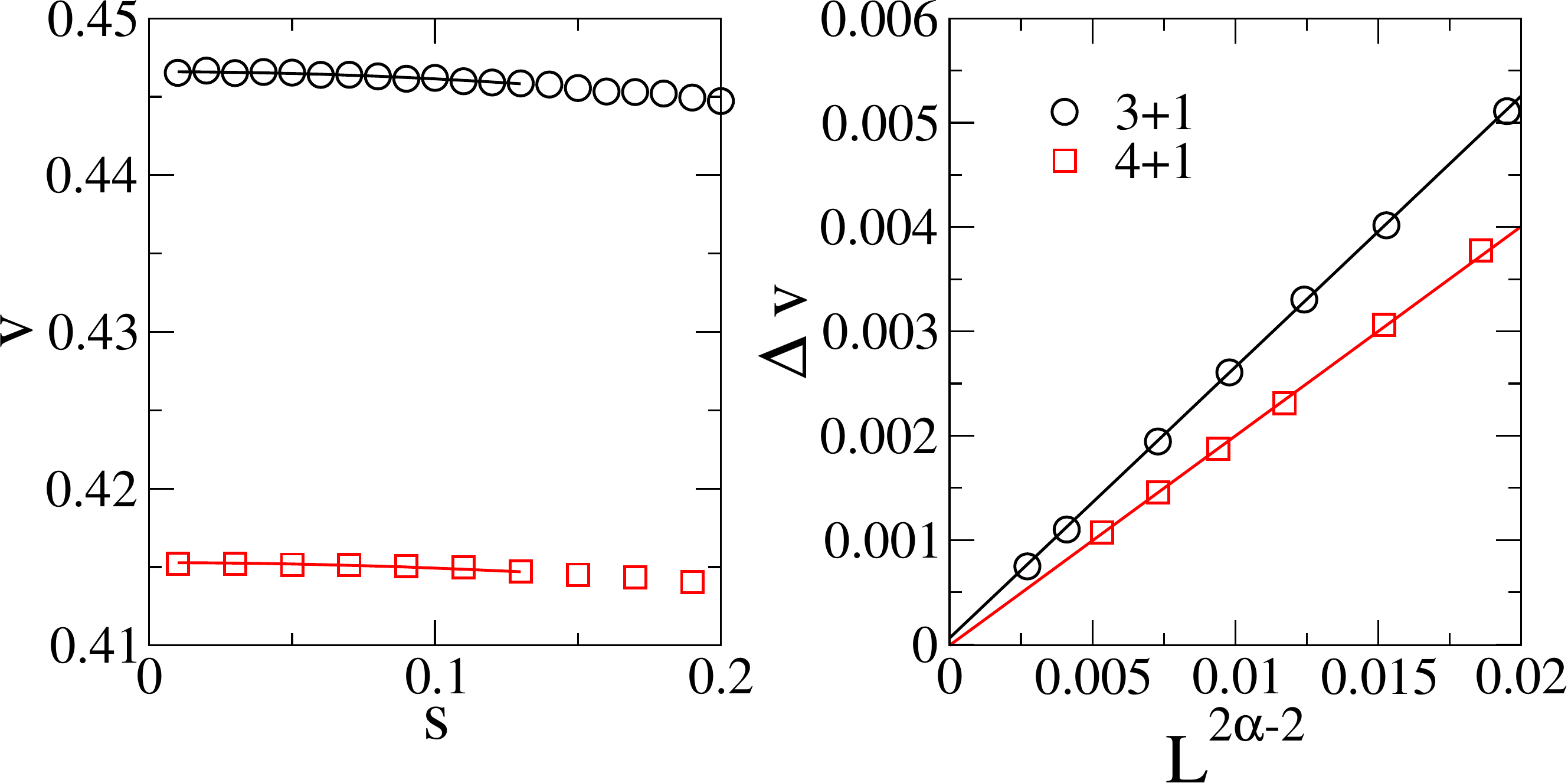}
\caption{\label{fig:par} Determination of non-universal parameters in RSOS model
with height restriction $m=2$ for three- and
four-dimensional substrates. Lines are parabolic (left) or linear (right)
regressions to determine the parameters $\lambda$ and $A$, respectively.}
\end{figure}

The cumulant ratio analysis allows us to obtain essentially all statistics of the
distribution in terms of the first cumulant~\cite{Alves13,Oliveira13}, which must
be determined independently using the non-universal
parameters $\Gamma$ and $v_\infty$~\cite{TakeuchiJSP12,Healy2}. The non-universal parameter controlling
the amplitude fluctuations in  KPZ ansatz can be obtained by the relation
$\Gamma=|\lambda| A^{1/\alpha}$ that,  apart from some dimensionless prefactor,
holds for KPZ equation below critical dimension ($\alpha, \beta >0$)~\cite{KrugPRA92}.
The parameter $\lambda$ can be determined using deposition on tilted
large substrates with an overall slope $s$, for which a simple dependence between
velocity and slope, $v=v_\infty+\frac{\lambda}{2}s^2$
is expected for the KPZ class~\cite{krug90}.
The parameter $A$ also has a simple relation with asymptotic velocity $v_L$ for
finite systems of size $L$~\cite{krug90},
$\Delta v = v_L-v_\infty \simeq -\frac{A\lambda}{2}L^{2\alpha-2}$.
We used the estimates of $\alpha$ exponents reported in Ref.~\cite{Odor2010}.
Figure~\ref{fig:par} shows the data analysis to determine $A$ and $\lambda$
parameters for $d=3$ and 4. We were not able to accurately perform these plots
for higher dimensions since tilting is hampered by the small
substrate sizes currently attainable. The values found are shown in
Table~\ref{tab:KPZpar}.

\begin{figure}[ht]
 \centering
 \includegraphics[width=8cm]{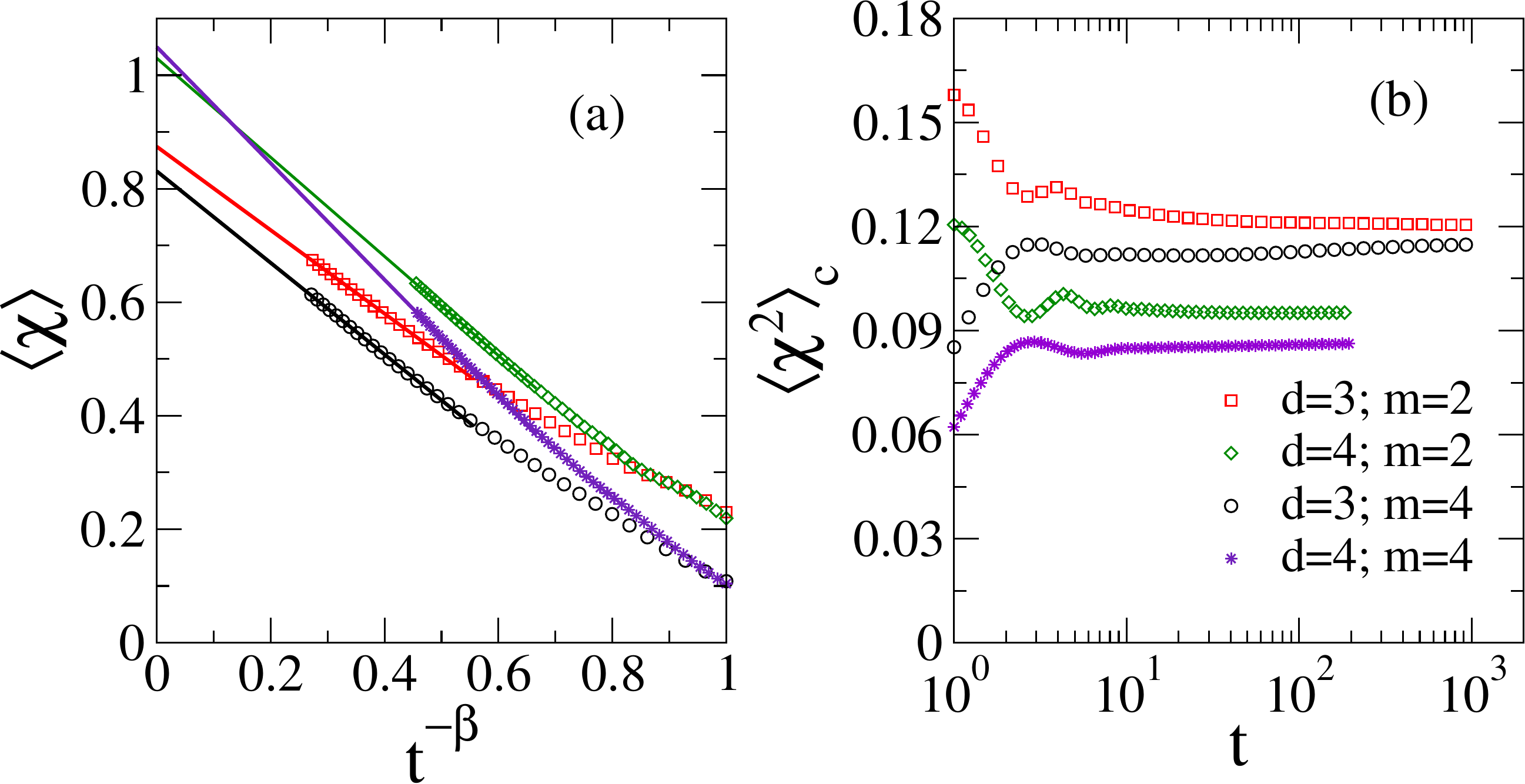}

 \includegraphics[width=8cm]{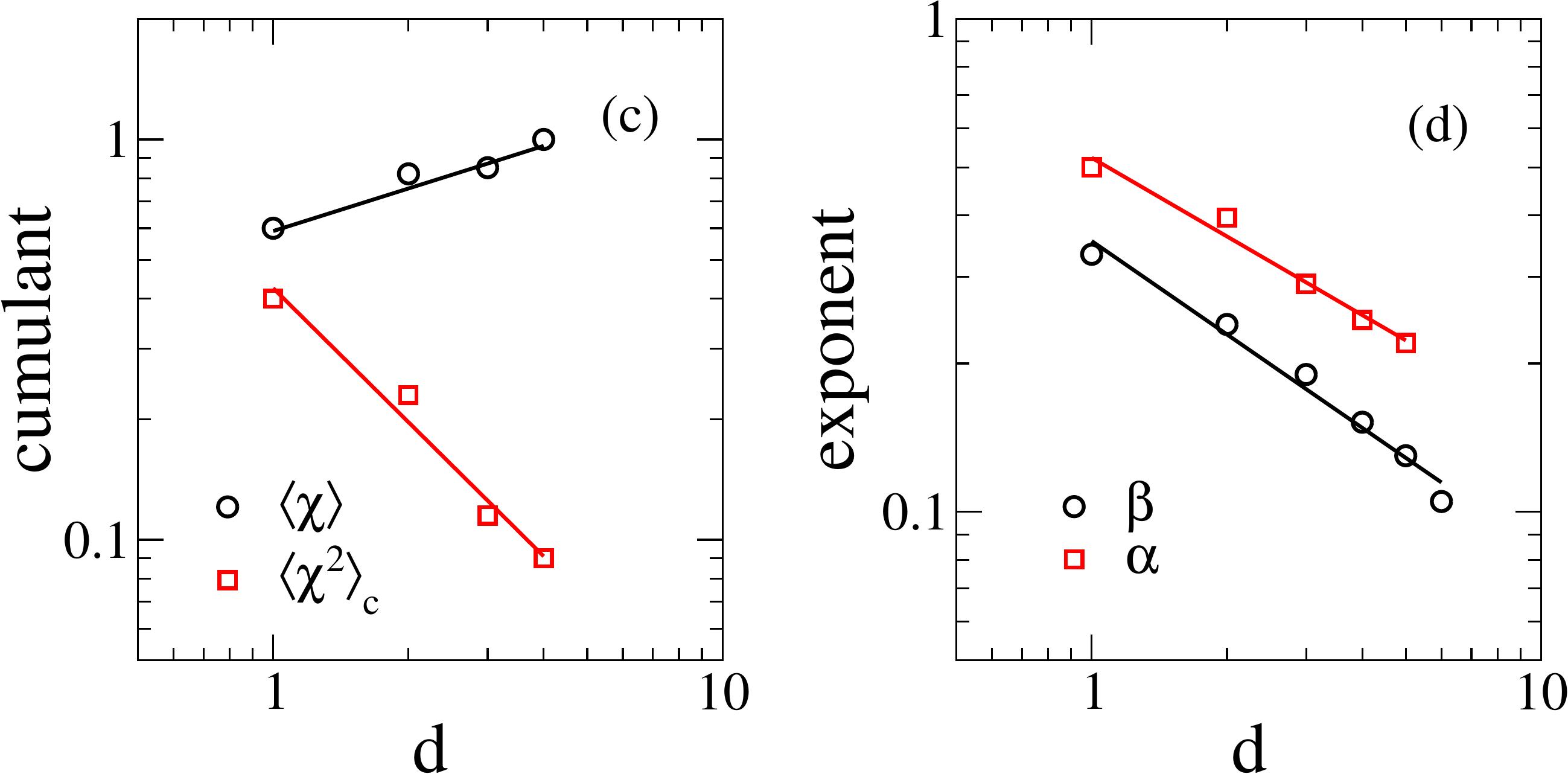}
 \caption{Determination of (a) first and (b) second cumulant of $\chi$. Lines are
linear regressions used to extrapolate $\lrangle{\chi}$. The dependence on the
substrate dimension of the  universal (c) cumulants and (d) exponents.
Lines are power law regressions.
The roughness exponents were taken from Ref.~\cite{Odor2010}.}
 \label{fig:chimed}
\end{figure}

\begin{figure}[ht]
 \centering
 \includegraphics[width=7cm]{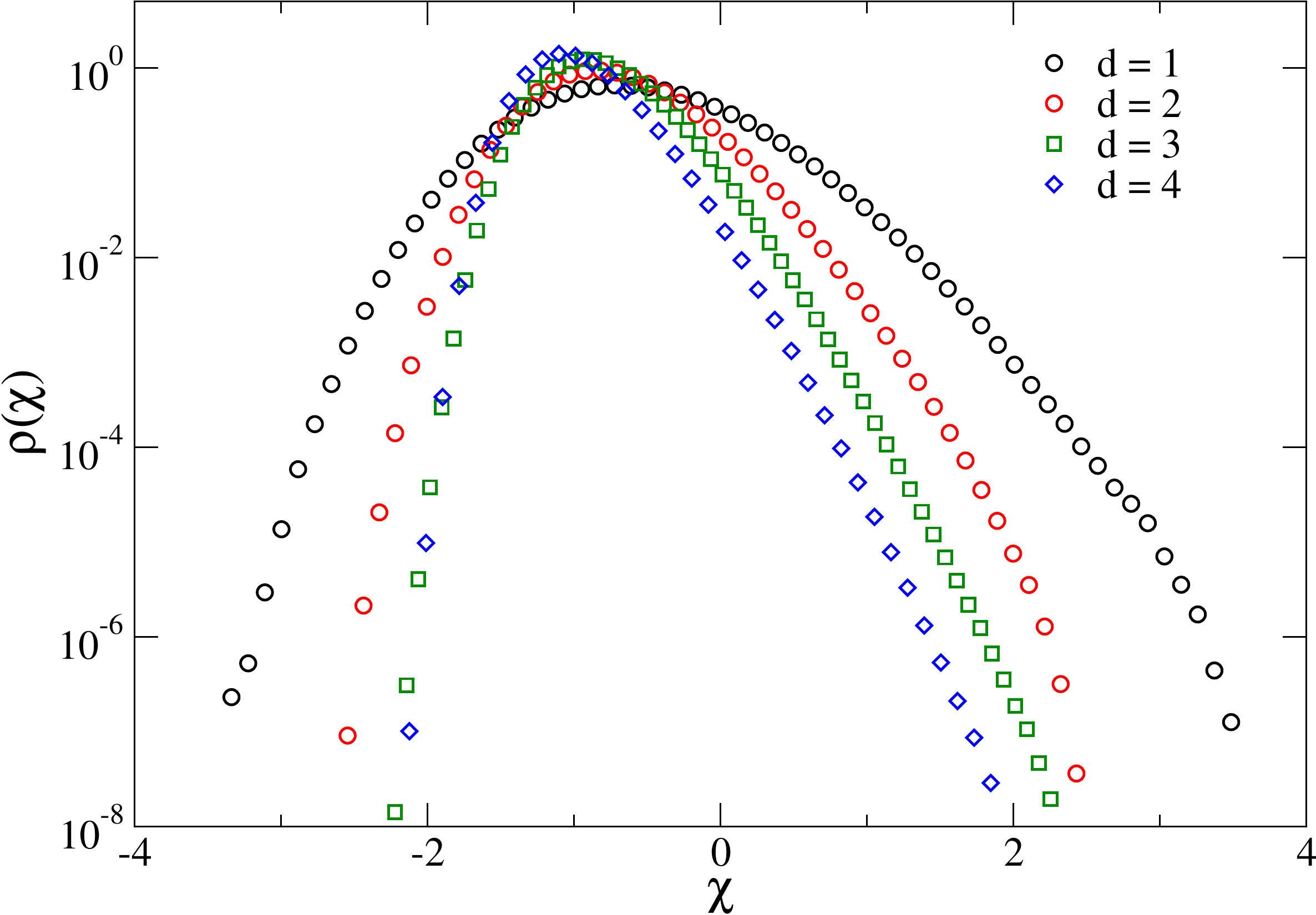}
 \caption{Density probability distribution for RSOS model at substrates
with dimension $d=1$ to $4$. The case
$d=1$ \textit{was  not} rescaled by the factor $2^{1/3}$ to render the Gaussian
orthogonal ensemble (GOE) distribution~\cite{TakeuchiJSP12}.}
 \label{fig:rhochi}
\end{figure}

The first cumulant can then be obtained using Eq.~(\ref{eq:htcorr}):
\begin{equation}
\lrangle{\chi}=\frac{\lrangle{h}-v_\infty t}{s_\lambda(\Gamma t)^{\beta}}
-\frac{\lrangle{\eta}}{s_\lambda(\Gamma t)^{\beta}}+\cdots,
\end{equation}
where the finite-time correction $t^{-\beta}$ is explicitly used to extrapolate
$\lrangle{\chi}$. Analogously, the second cumulant can be obtained using
$\lrangle{\chi^2}_c\simeq \lrangle{h^2}_c/(\Gamma t)^{2\beta}+\cdots$, in which
corrections do not follow a universal
scheme~\cite{Oliveira12,Alves13,TakeuchiJSP12,Oliveira13}. Figures~\ref{fig:chimed}(a) and
(b) show typical plots for cumulant determination. As one can notice, the
extrapolation is imperative to the estimate of $\lrangle{\chi}$ {from finite-time data}, fact neglected
in {the first analysis of two-dimensional models at light on the KPZ ansatz~\cite{Healy}}. Indeed, our
estimate of $\lrangle{\chi}$ for $d=2$ is slightly larger than the former
estimate for RSOS model~\cite{Healy} {but completely consistent with a more
refined analysis done later~\cite{Healy2}}. The asymptotic cumulants are shown in
Table~\ref{tab:KPZpar}.  One can observed  that the first cumulant gets more
negative,  while the variance decreases  as dimension increases. Finally,
possessing $v_\infty$, $\Gamma$ and $\left\langle \eta \right\rangle $ the
density probability distribution can  be drawn for different
dimensions, see Fig.~\ref{fig:rhochi}.
What do we observe is that distributions vary less as dimension
is increased but no special hallmark can be highlighted.

Further evidences supporting the absence of an upper critical dimension
are given in Figs.~\ref{fig:chimed}(c) and (d) where cumulants
and exponents are drawn against substrate dimension. In both pictures
no signature of an upper critical behavior can be resolved. On the contrary,
both scaling exponents and variance seem to slowly converge  to zero roughly obeying
power laws $\beta\sim d^{-0.63}$, $\alpha\sim d^{-0.53}$, and
$\lrangle{\chi^2}_c\sim d^{-1.1}$. In particular, the  RG analysis of
Castellano \textit{et al}.~\cite{Castellano1998b} foresaw
a roughness exponent decaying slower than $1/d$ that is
fully supported by our current analysis.

\begin{table}[ht]
\caption{\label{tab:KPZpar} Estimates of non-universal parameters ($A$,
$\lambda$, $\Gamma$) for RSOS model in $d=1$ to 4 dimensions.
Height restriction parameters are shown in  brackets. The estimates of the
first and second cumulant of  $\chi$ are  shown in
last columns.
Results for $d=1$  were extracted from Ref.~\cite{Oliveira12} where a factor
different convention $\Gamma=|\lambda| A/2$ was used.  {Our results in $d=1$ and
2 with $m=1$ are in agreement with former reports~\cite{Healy2,KrugPRA92}.}}

\begin{tabular}{cccccc} \hline\hline
  $d$ [$m$] & $A$     & $\lambda$ & $\Gamma$& $\lrangle{\chi}$ & $\lrangle{\chi^2}_c$ \\ \hline
  1 [1]~& 0.81          & -0.77        &  0.51   &   -0.60        & 0.40\\
 2 [1]  & 1.22(4) &    -0.41(1) & 0.68(6)  &-0.83(2)    & 0.23(1)\\
 2 [2]  & 4.5(1)  &    -0.121(3) & 5.5(2)  &-0.82(2)    & 0.23(1)\\
 3 [2] &5.8(2)  & -0.090(2)    &  38(3)         &-0.86(2)       & 0.12(1)\\
 3 [4] & 19(2)   & -0.024(2)   &  600(50)        & -0.82(3)      & 0.11(1) \\
 4 [2] & 8(1) & -0.05(1)        &  240(50)        &-1.00(4)       & 0.09(1)\\
 4 [4] & 25(2) & -0.015(2)      &  7600(900)       &-0.98(5)       & 0.09(1)\\ \hline \hline
 \end{tabular}
\end{table}

In summary, we performed extensive simulations of the RSOS model on substrates
with dimension up to $d=6$. We showed that the KPZ ansatz, given by
Eq.~(\ref{eq:htcorr}) and initially conjectured for $d=1$, holds also in
dimensions $d=3-6$, extending a recent generalization to
$d=2$~\cite{Oliveira12,Healy}. Furthermore, the asymptotic growth velocities
were shown to follow the slope and size dependence predicted by the theoretical
machinery for the KPZ class~\cite{krug90}.
The height distributions found are independent of the height restriction
parameter, pointing out its universality. Our results also rule out a critical
dimension $d_u\le 6$. The extrapolations of universal quantities to
higher dimensions are consistent with the absence of an upper critical dimension.
{We expect that our results will motivate the analysis of curved
growth in high dimensions, extending the geometry-dependent universality of
$\chi$ to $d\ge3$. Additionally,  the  high-dimensional
analogues of the Airy processes for spatial covariance in $d=1$~\cite{krugrev},
which only very recently was numerically determined in $d=2$~\cite{Healy2014},
is an interesting problem that deserves further attention.}

\begin{acknowledgments}
Authors acknowledge the support from CNPq and FAPEMIG (Brazilian agencies).
{Authors thanks Timothy Halpin-Healy and Zolt\'an R\'acz for suggestions
and criticisms on the manuscript.}
\end{acknowledgments}


%

\end{document}